\documentclass[preprint]{aastex61}
\usepackage{CJK}

\newcommand\aastex{AAS\TeX}

\newcommand\um{$\mu \textrm{m}$}
\newcommand\mjybeam{mJy beam$^{-1}$}
\newcommand\jybeam{Jy beam$^{-1}$}
\newcommand\kms{km s$^{-1}$}

\received{...}
\revised{...}
\accepted{...}
\shorttitle{\aastex\ Magnetic fields in L1448 IRS 2}
\shortauthors{Kwon et al.}
\begin{document}

\title{Highly Ordered and Pinched Magnetic Fields in the 
Class 0 Proto-binary System L1448 IRS 2}

\correspondingauthor{Woojin Kwon}
\email{wkwon@kasi.re.kr}

\author{Woojin Kwon}
\affiliation{Korea Astronomy and Space Science Institute (KASI),
776 Daedeokdae-ro, Yuseong-gu, Daejeon 34055, Republic of Korea}
\affiliation{University of Science and Technology, Korea (UST), 
217 Gajeong-ro, Yuseong-gu, Daejeon 34113, Republic of Korea}

\author{Ian W. Stephens}
\affiliation{Harvard-Smithsonian Center for Astrophysics,
60 Garden Street, Cambridge, MA, USA}

\author{John J. Tobin}
\affiliation{National Radio Astronomy Observatory, 520 Edgemont Rd.,
Charlottesville, VA 22903, USA}
\affiliation{Homer L. Dodge Department of Physics and Astronomy, 
University of Oklahoma, 440 W. Brooks Street, Norman, Oklahoma 73019, USA}
\affiliation{Leiden Observatory, Leiden University, P.O. Box 9513, 2300-RA Leiden, 
The Netherlands}

\author{Leslie W. Looney}
\affiliation{Department of Astronomy, University of Illinois, 
1002 West Green Street, Urbana, IL 61801, USA}

\author{Zhi-Yun Li}
\affiliation{Astronomy Department, University of Virginia, 
Charlottesville, VA 22904, USA}

\author{Floris F. S. van der Tak}
\affiliation{SRON Netherlands Institute for Space Research, 
Landleven 12, 9747 AD Groningen, The Netherlands}
\affiliation{Kapteyn Astronomical Institute, University of Groningen, 
PO Box 800, 9700 AV Groningen, The Netherlands}

\author{Richard M. Crutcher}
\affiliation{Department of Astronomy, University of Illinois, 
1002 West Green Street, Urbana, IL 61801, USA}



\begin{abstract}

We have carried out polarimetric observations with the Atacama Large
Millimeter/submillimeter Array (ALMA) toward the Class 0 protostellar
system L1448 IRS 2, which is a proto-binary embedded within a
flattened, rotating structure, and for which a hint of a central
disk has been suggested, but whose magnetic fields are aligned with
the bipolar outflow on the cloud core scale.  Our high sensitivity {and high resolution
($\sim 100$ au)} observations show a {clear} hourglass
{magnetic field morphology centered on} the protostellar system,
but the central pattern is consistent with a toroidal field
indicative of a circumstellar disk, {although other interpretations are 
also possible, including field lines dragged by an equatorial accretion flow 
into a configuration parallel to the midplane.} {If a relatively large
disk does exist, it would suggest that the magnetic braking catastrophe
is averted in this system, not through a large misalignment between
the magnetic and rotation axes, but rather through some other
mechanisms, such as non-ideal magneto-hydrodynamic effects and/or
turbulence.} We have also found a relationship of decreasing
polarization fractions with intensities {and the various slopes
of this relationship can be understood as multiple polarization
mechanisms and/or depolarization from a
changing field morphology.} In addition, we found {a prominent
clumpy depolarization strip} crossing the center perpendicular to
the bipolar outflow.  Moreover, {a rough estimate of the magnetic
field strength indicates that the field is} strong enough to hinder
formation of a rotationally supported disk, which is inconsistent
with the feature of a central toroidal field. 

\end{abstract}




\section{Introduction} \label{sec:intro}

Magnetic fields are thought to play a significant role in star
formation on all scales from the cloud ($\sim 1$ pc) to 
the disk ($\sim$ 100 au).  For example, it
has been found that magnetic field directions are well ordered and
typically perpendicular to parsec-scale filamentary structures
\citep[e.g.,][]{2013A&A...550A..38P},
which indicates that magnetic fields are important
to form such intermediate scale structures. Also, the magnetic
energy is comparable to the kinetic energy down to a few thousand
au scales \citep[e.g.,][]{2014arXiv1404.2024L,2017ApJ...846..122P}.

In addition, magnetic fields can affect circumstellar disk formation
at the early protostellar stages. 
\citet{2013ApJ...768..159H} found
that magnetic field directions of 16 young stellar objects (YSOs) are
rather random with respect to their bipolar outflows on a few hundred au
scales, although a couple of examples with hourglass morphology magnetic 
fields aligned to its bipolar outflow had been known at the time 
\citep{2006Sci...313..812G,2013ApJ...769L..15S}.
Later, it has
been suggested that the magnetic field directions of YSOs can be
understood with the existence of an extended disk structure
at the youngest YSOs, the so-called Class 0 YSOs 
\citep[e.g.,][]{2015ApJ...798L...2S}.
For example, L1527 has a magnetic field morphology
perpendicular to the bipolar outflow and has an extended Keplerian disk 
(radius $\sim$ 54 au) 
\citep[][]{2012Natur.492...83T, 2013ApJ...768..159H, 2014ApJ...796..131O},
whereas
L1157 that has a magnetic field aligned with the bipolar outflow has
no disk structure larger than 15 au \citep[][]{2013ApJ...769L..15S, 
2013ApJ...779...93T}.
Such features have been explained by magnetic braking, which
can be so efficient in YSOs having a magnetic field aligned with the bipolar
outflow that a rotation-supported disk structure is largely suppressed
at the early stages 
\citep[e.g.,][]{2008ApJ...681.1356M,Hennebelle:2009bi,2012A&A...543A.128J,2018MNRAS.tmp..552M}, 
often called the magnetic braking catastrophe of disk formation.

However, not all Class 0 YSOs fit the interpretation connecting
the magnetic field morphology and the disk structure. For example,
although a rotating disk-like structure has been detected
around L1448 IRS 2 and its central binary companion 
\citep[e.g.,][]{2015ApJ...805..125T, 2016ApJ...818...73T},
the magnetic field
direction detected on 500--1000 au scales is mostly aligned with 
the bipolar outflow \citep{2014ApJS..213...13H},
in which magnetic braking is expected to be efficient.
In addition, \citet{2014ApJ...797...74D} reported that
the most preferred magnetic field for the Class 0 YSO L1527, which
has a large Keplerian disk \citep{2012Natur.492...83T, 2014ApJ...796..131O},
is a weak field aligned with the bipolar outflow, not a perpendicular
field, when considering magnetic fields of large $\sim3000$ au scales 
as well as small $\sim500$ au scales.
On the other hand, as an example,
\citet{2014MNRAS.438.2278M} reported that the magnetic braking
effect depends on density distributions of the dense cores that collapse
to form the YSOs as well as magnetic
field morphologies. Also, it has been discussed that
numerical simulations using a large sink radius suppress disk formation
at the early evolutionary stages. In addition, non-ideal magneto-hydrodynamic (MHD)
effects (e.g., Ohmic dissipation, {ambipolar diffusion, and Hall effect}) 
can enable the formation of a {small} rotationally supported disk even in the case
of a magnetic field aligned with the bipolar outflow that has efficient
braking in the ideal MHD simulations 
{\citep[e.g.,][]{2010ApJ...718L..58I,2011ApJ...733...54K, Dapp:2012fo, 
2015ApJ...801..117T, {2015ApJ...810L..26T}, 2018MNRAS.473.4868Z}.}

An obvious way to investigate whether primordial magnetic field 
morphologies affect disk formation at the early evolutionary stages 
is to examine the small scale fields of those YSOs that have envelope-scale
(1000--10000 au) fields aligned with the bipolar outflow. 
In this paper, we report polarimetric observations of the
Atacama Large Millimeter/submillimeter Array (ALMA) toward L1448 IRS 2
focusing on how the magnetic fields change on 100 au scales.

\section{Target and Observations} \label{sec:obs}

Several Class 0 YSOs with flattened envelope structures have been
identified by \citet{2010ApJ...712.1010T} in 
the Perseus molecular cloud at a distance of about 300 pc
\citep[][]{2018arXiv180308931Z, OrtizLeon:2018wz}.
Of these envelopes,
L1448 IRS 2 has the clearest flattened structure 
and has been observed in polarization indicating a magnetic
field aligned with the bipolar outflow \citep{2014ApJS..213...13H}.
Previous studies have imaged a large, extended bipolar outflow 
originating from the target at various wavelengths:
e.g., near IR observations 
using the {\em Spitzer Space Telescope} \citep{2007ApJ...659.1404T}.
In addition, \citet{2009ApJ...696..841K} 
reported that grains have significantly
grown based on the dust opacity spectral index estimated
from 1 and 3 mm observations of
the Combined Array for Research in Millimeter-wave Astronomy 
(CARMA).
Regarding polarimetric observations, Caltech Submillimeter
Observatory SHARP observations detected $350$ \um\ continuum polarization
perpendicular (inferred magnetic field\footnote{
Assuming magnetic field grain alignment, the magnetic field directions 
are inferred by $90\degr$ rotation of polarization
directions.} parallel) 
to the bipolar outflow with a $10 \arcsec$ angular resolution
\citep{2013ApJ...770..151C}.
\citet{2014ApJS..213...13H} also detected polarization in the
same direction, particularly in the blueshifted lobe on the northwest 
side from the center with an angular resolution of $\sim2 \arcsec$.

Polarimetric observations toward L1448 IRS 2 in ALMA Band 6 were made
on 2016 November 12 and 14 (2016.1.00604.S, PI: Woojin Kwon). 
Individual tracks were run over 3 hours
to achieve a good parallactic angle coverage.
HH 211 was observed simultaneously with L1448 IRS 2 and shared 
the same phase calibrator; HH 211 will be reported in a separate paper.
The November 12 and 14 tracks have 42 and 40 antennas in the
array, respectively.
J0238+1636 was used as polarization calibrator and flux calibrator.
Its flux was set to 1.085 Jy at 233 GHz with a
spectral index of about -0.45.
J0237+2848 and J0336+3218 were bandpass and phase calibrators,
respectively.
All the execution blocks were calibrated separately and
combined when making images.
Images were made using a Brigg's weighting with a robust 
parameter of 0.5. We found this weighting was a good compromise 
between resolution and sensitivity.
The final image has a synthesized beam
of $0\farcs57\times0\farcs37$ (PA = $9.14\degr$).
The noise levels of Stokes I (total intensity), Q, and U maps
are $\sim0.10$, $\sim0.014$, and $\sim0.014$ \mjybeam, respectively.
{The polarization intensity map achieved by the Stokes Q and U 
maps with a debias using the Stokes Q and U map noise level has a RMS 
noise of 0.008 \mjybeam.} 

\section{Results} \label{sec:results}

\subsection{Magnetic field morphology} \label{subsec:tables}

\begin{figure}
\plottwo{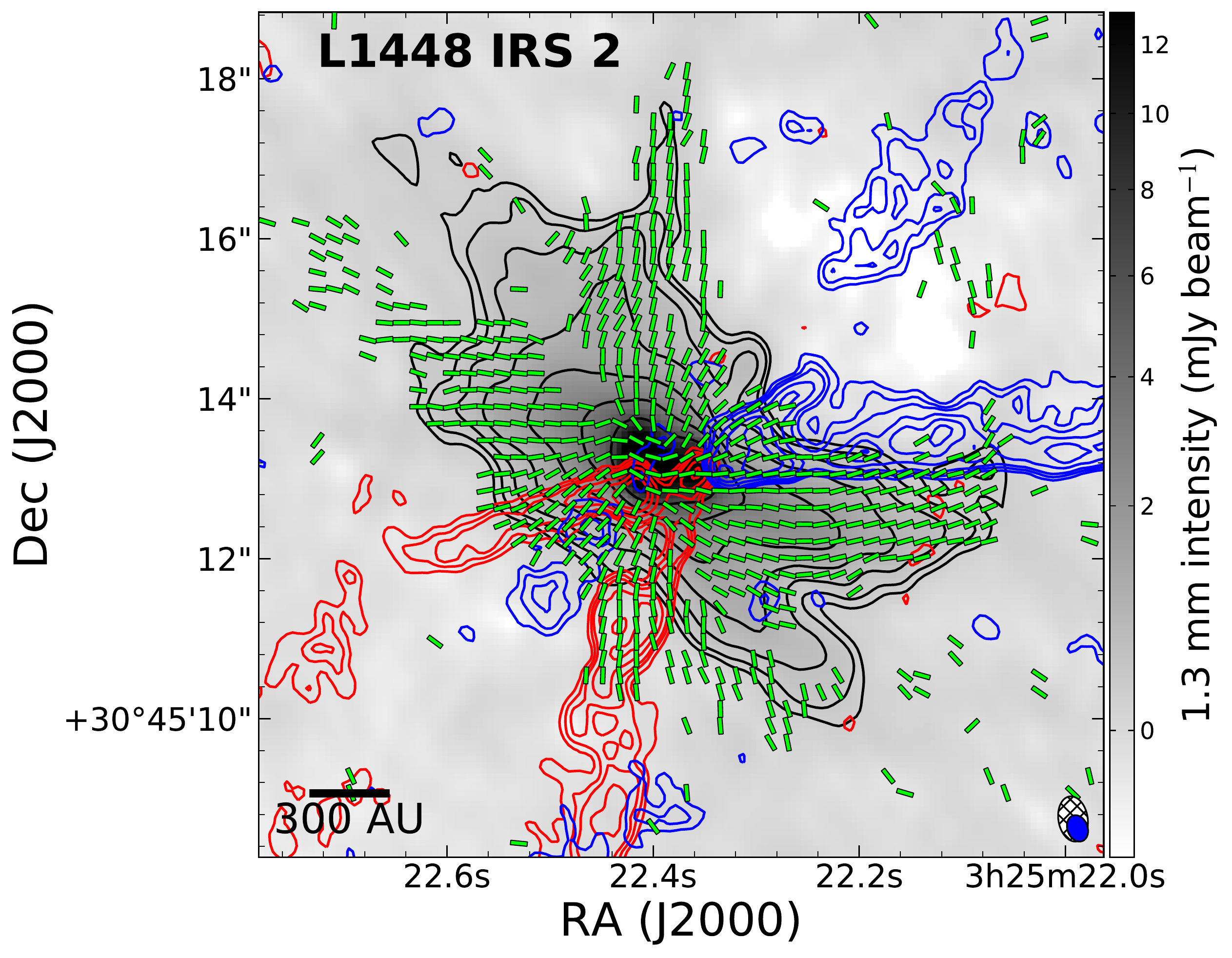}{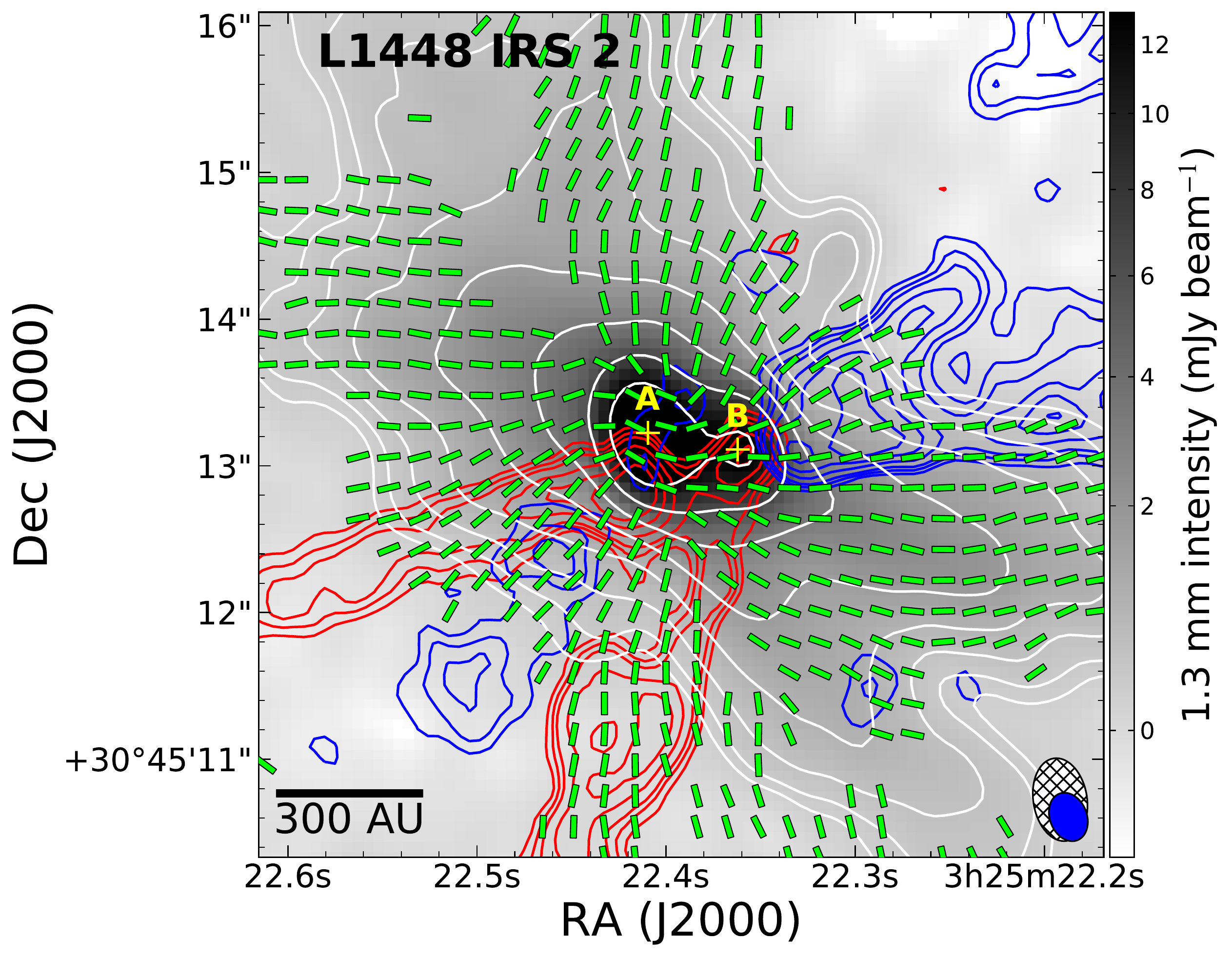}
\caption{Magnetic field morphology around L1448 IRS 2. The green vectors
have been rotated by $90\degr$ from the polarization directions
to indicate the inferred magnetic field direction. 
{Vectors of $2\sigma$ level or better detections are marked every 
$0.21\arcsec$, which is comparable to the Nyquist sampling.}
The gray scale and black and white contours present the total intensity 
distribution with levels of 2, 3, 5, 9, 17, 33, 65, and 129 times 0.1
\mjybeam. The blue and red contours are CO 2-1 intensity distributions
integrated in velocity ranges of $-8.0$ to 2.0 and 7.5 to 16.0 \kms, respectively,
at levels of 3, 4, 5, 7, and 9 times 11 K \kms. The central
region zoomed-in is presented in the right. The synthesized beams
of the CO and the continuum data are marked in the bottom right corner
in blue and white, respectively. 
The binary system positions are 2A(03:25:22.407 +30:45:13.21) and
2B(03:25:22.363 +30:45:13.12).}
\label{fig_mag_out}
\end{figure}

{Magnetic fields are inferred in perpendicular
to polarizations of dust thermal emission in millimeter/submillimeter
wavelengths, because non-spherical dust grains are aligned with
their minor axis (spin axis) parallel to magnetic fields 
\citep[e.g.,][]{{2007MNRAS.378..910L}}.
In Figure \ref{fig_mag_out}, we rotate the polarization directions 
by 90 degree and infer the magnetic fields.} 
The magnetic field morphology shows a beautifully {clear} hourglass 
morphology perpendicular to the elongated structure.
Consistent with previous CARMA observations 
\citep{2014ApJS..213...13H}, the region northwest of the center shows
a poloidal field, which is approximately aligned with the bipolar outflow
direction. However, in the central region the field direction rapidly 
changes to an orientation perpendicular to the bipolar outflow, which
is consistent with the polarization pattern produced by grains 
aligned with a toroidal magnetic field close to the protostar. 

{Recently, however, it has also been reported that magnetic fields
on disk scales
cannot be inferred from polarization patterns 
because other polarization mechanisms may be dominant. 
\citet{2017ApJ...851...55S} 
showed that polarization of
dust thermal emission in HL Tau dramatically changes with wavelengths.
In Band 3 (3 mm) the polarization pattern is azimuthal, which may be understood
by anisotropic radiation alignment \citep{{2017ApJ...839...56T},{2019MNRAS.483.2371Y}}, 
while in Band 7 (0.85 mm) it is parallel
to the disk minor axis, which is indicative of scattering. 
The intermediate wavelength, Band 6 (1.3 mm),
the pattern is a combination of the two. These results 
indicate that polarization may not be able to probe magnetic fields.
In contrast, \citet{2018A&A...616A..56A}
found that the Class I YSO BHB07-11 has a uniform polarization
pattern over Band 3, 6, and 7 of ALMA. This suggests that dust
grains are not large, and that they are aligned with magnetic fields.
On the other hand, \citet{{2018A&A...617A...3L}} addressed that
polarizations can be caused by extinction even in millimeter
wavelengths, particularly toward high optical depth regions having
temperature increasing with distance from the observer along the line of sight. 
In such cases, polarization
fractions could be high (several percent or larger depending on
optical depth and background and foreground temperatures), and
magnetic fields are inferred parallel to the polarization direction,
unlike the usual case of polarizations in the thermal dust emission
of (sub)millimeter wavelengths.

With our data set taken in Band 6 toward the Class 0 YSO L1448 IRS
2, which is discussed in this paper, we argue that the polarizations
come from dust grains aligned with magnetic fields.  Regarding the
central region, other polarization mechanisms are worth discussing.
First, based on the polarization pattern at the center, with E-vectors
parallel to the bipolar outflow, it is unlikely caused by alignment
of anisotropic radiation.  Also, the intensity peak is about 26
\mjybeam, which corresponds to an optical depth $\tau \sim 0.2$
when assuming $T = 30$ K, so the polarization may not be caused by
extinction.  Therefore, the magnetic fields may be inferred by 90
degree rotation of polarization directions, 
which results in a pattern in the central region that is broadly
consistent with a toroidal magnetic field projected in the sky
plane.
On the other hand, self-scattering cannot be ruled
out by the data set taken only in Band 6.  However, note that even
in this case, it may provide indirect evidence for the presence of a circumstellar disk(-like)
structure at the center, since self-scattering has been found so
far only toward disks with large grains. Therefore, the main result of
this paper does not change. The ambiguity will be tackled by further
polarimetric observations in different wavelengths.  Indeed, we
have carried out polarimetric observations in ALMA Band 3 as well,
and the preliminary results support the interpretation of magnetic
fields.  These results will be discussed in a following paper (Kwon
et al. in prep).
}

{In addition, 
there is the possibility that the central polarozation pattern is
produced by magnetic field lines that are dragged by an equatorial
accretion flow into a configuration that is parallel to the midplane
of the system. However, we believe this possibility is less likely
than the rotationally-induced toroidal field interpretation because
there is already evidence for significant velocity gradient along
the equatorial plane \citep[see, e.g. Fig. 11 of][]{2018ApJ...867...43T}.
Nevertheless, higher resolution line observations are needed to
firmly establish  whether the intriguing field orientation near the
center is produced by accretion or rotation.}
{Note the seven vectors around the source A within or on the
inner most contour in the right zoomed-in plot of Figure \ref{fig_mag_out}.  
A detailed modeling
may be required to constrain the structure size of a toroidal
magnetic field. However, based on the number of vectors showing the
shifted direction, which is 1.5 of the beam area, the central disk
could be up to about 50 au in radius: {considering a beam smoothing,
the real structure would be about 0.5 of the beam area, so $r = (0.5 A_{beam}/\pi)^{0.5}d$,
where $A_{beam}$ is the beam area and $d$ is the target distance}. 
Further molecular line
observations at a high angular resolution will provide the info on
how large the disk is, if it existed, and whether it is rotationally
supported.}

Note that the self-scattering polarization pattern is expected to be
parallel to the minor axis of an inclined disk
\citep[e.g.,][]{Yang:2015kg,2017ApJ...844L...5K,2017ApJ...851...55S},
which means that after $90\degr$ rotation, the corresponding B vectors
look like a toroidal feature.
If the toroidal field interpretation is correct, it would indicate
that rotation has become fast and energetic enough  to wind up the
field lines, which likely signals the formation of a rapidly rotating
disk. If the dust self-scattering interpretation is correct, it
would indicate that grains in the flattened structure on the scale
of several tens au have grown to roughly 100~$\mu$m sizes or more, which
again would favor the existence of a rotationally supported disk that
is conducive to grain growth through a higher density and longer
time compared to a dynamically collapsing inner envelope.
Recently, some other ALMA polarimetric observations have also presented 
polarization patterns of self-scattering or toroidal magnetic fields 
in the central regions of Class 0 and I YSOs with a disk 
\citep{2018ApJ...854...56L,2018ApJ...859..165S}
and disk candidates \citep{2018ApJ...855...92C}.
{Additional polarimetric observations at different wavelengths will 
allow us to distinguish magnetic field alignment from self-scattering
\citep[e.g.,][, Kwon et al. in prep]{2018A&A...616A..56A}.}

The contours in the right panel of Figure \ref{fig_mag_out} also 
show the binary companion (L1448 IRS 2B),
which is separated from the primary by about $0.6\arcsec$ 
(corresponding to $\sim180$ au
at the target distance) toward the west. This companion is less
bright in the 1 mm continuum and has been detected at 9 mm (Ka-band) 
\citep{2016ApJ...818...73T} and at 1 mm \citep{2018ApJ...867...43T}.
The blue and red contours overlaid in the figure are
CO 2-1 molecular line data taken by ALMA 
\citep[2013.1.00031.S;][]{2018ApJ...867...43T}.
The angular resolution of the CO 2-1 observations is 
$0\farcs35 \times 0\farcs25$ (PA: $21\degr$),
which is slightly better than the polarimetric continuum data.
Since these observations lack short baselines, only the cavity walls 
were detected, as the extended features between the walls 
\citep[as detected in][]{2015ApJ...805..125T} are filtered out.
Interestingly, the less bright companion L1448 IRS 2B 
seems to be more coincident with the bipolar outflow.
While the redshifted lobe is primarily centered on the combination of
both L1448 IRS 2A and 2B, the main blueshifted lobe seems to be centered on
L1448 IRS 2B only.
However, there is a weak blueshifted
feature from the L1448 IRS 2A and along the left continuum branch.
It is possible that the blueshifted component from L1448 IRS 2A might
be in the same velocity regime of the ambient cloud and thus could be 
filtered out by the interferometer.

\subsection{Polarization intensity and fraction} \label{subsec:pi}

\begin{figure}
\includegraphics[scale=0.28]{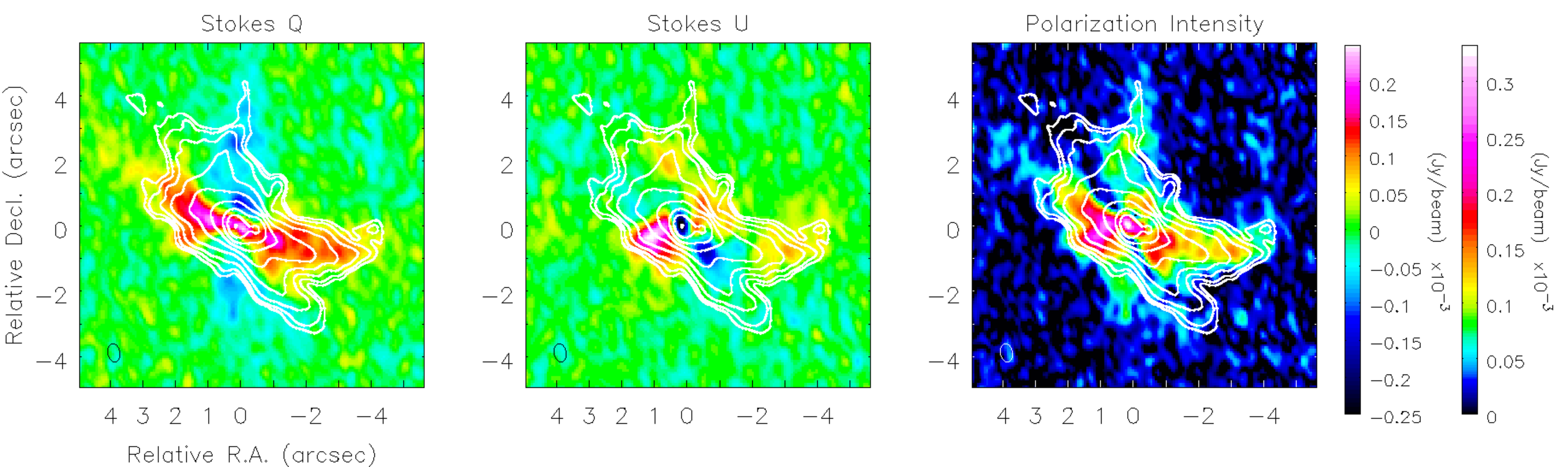}
\caption{Stokes Q, Stokes U, and polarization intensity maps are
on left, in the middle, and on right, respectively.
The contours present distributions of total intensity (Stokes I)
at the same levels of Fig. \ref{fig_mag_out}. {The color edges
are for the Stokes Q and U maps (the left covering minus and plus values) 
and the polarization intensity map (the right).}
\label{fig_iqu}}
\end{figure}

\begin{figure} 
\begin{tabular}{cc}
\includegraphics[scale=0.45]{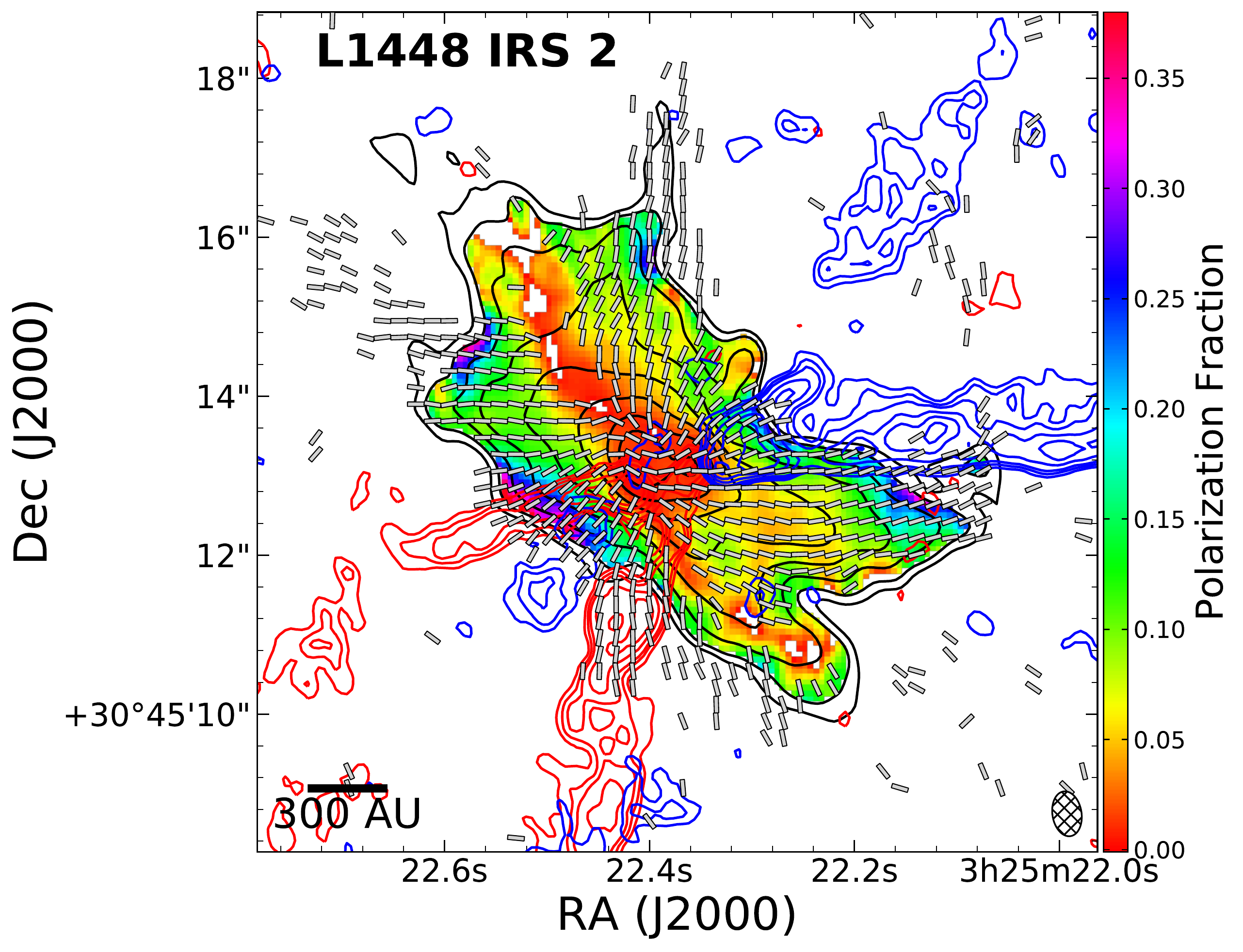} &
\includegraphics[scale=0.30]{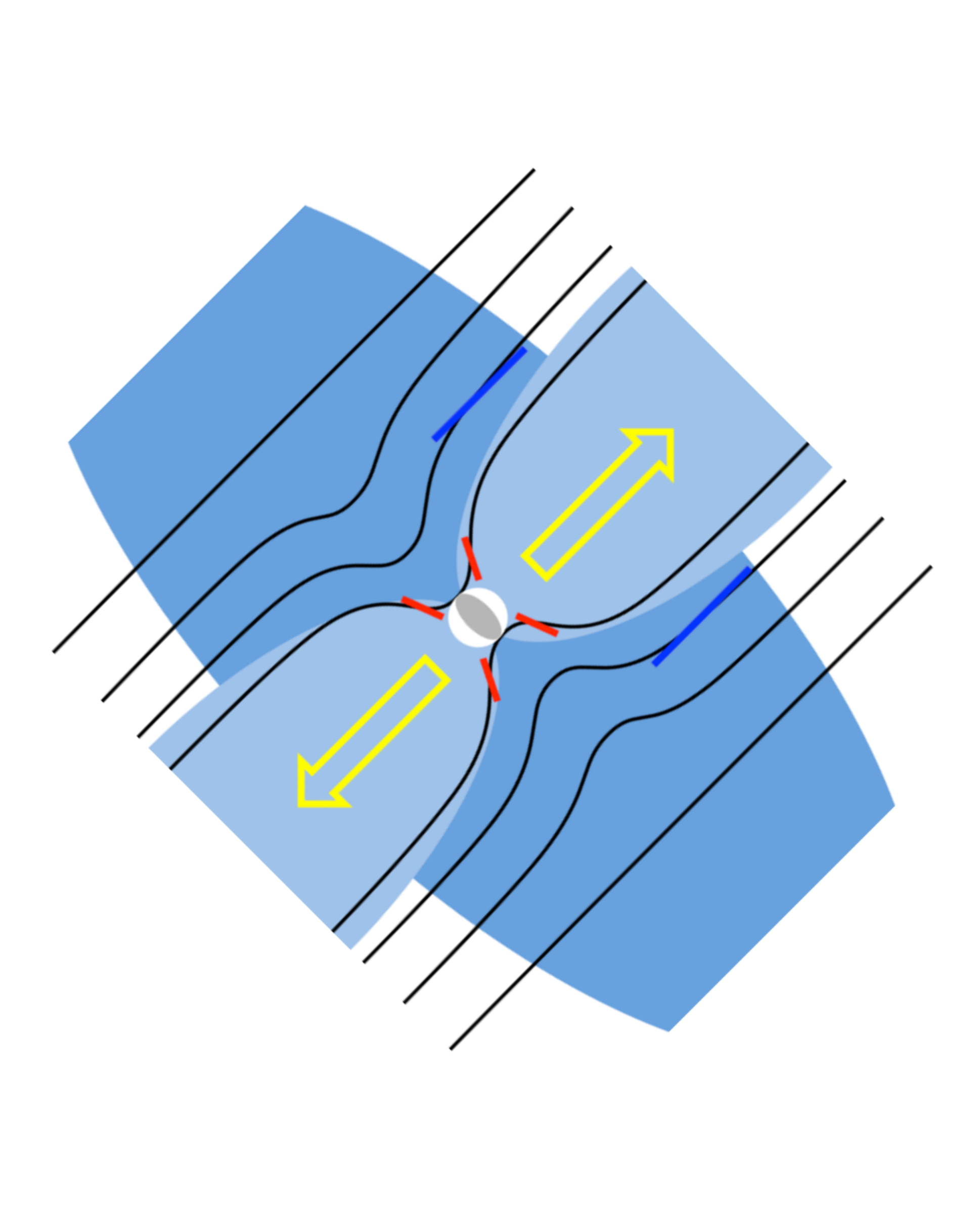}\\
\end{tabular}
\caption{Polarization
fraction distribution indicated in color scales. The other plot
components are the same as Fig. \ref{fig_mag_out}. The schematic diagram
on the right illustrates hourglass morphology magnetic fields detected in
different alignment: small scales in red and large scales in blue.
\label{fig_polratio}}
\end{figure}

In Figure \ref{fig_iqu}, 
the outflow cavity walls have relatively high polarization intensities,
while there is a strip across the center, almost perpendicular (P.A. 
$\sim 35\degr$) to the bipolar outflow 
\citep[P.A. $\sim 118\degr$ at large scales;][]{2017ApJ...846...16S}, 
with weak polarization signal lower than a few tens $\mu$\jybeam.
This depolarized strip is shown in more detail in Figure
\ref{fig_polratio} and discussed below.
The polarization intensity is not symmetrically distributed.
The central region of the total intensity peak has the highest
polarization intensity. Also, the region south of the 
depolarized strip to the east of the center 
is high in polarization intensity. 
In addition, polarization intensity is very clumpy 
compared to the total intensity distributions, which is
indicative that polarization is significantly affected by the local
environment.

As shown in Figure \ref{fig_polratio},
the depolarization regions of L1448 IRS 2 clearly appear at the
central region and along the strip perpendicular to the bipolar
outflow, whose polarization fractions are only a few percent or 
less.  The central
region with the binary system presumably has the most complicated
magnetic fields. Also, as addressed, the magnetic fields are changing
from aligned to perpendicular with respect to the bipolar outflow going from
large to small scales. These complicated polarization patterns
that are smaller than the beam reduce the measured polarization
fraction: i.e., beam smearing.
{A complicated magnetic field could also be caused by turbulence.
However, in the case of the turbulence-induced, depolarized regions
are rather randomly distributed \citep{2017ApJ...834..201L}.}

On the other hand, the depolarized strip is similar to the case 
of {an {\em inclined} cloud with a hourglass-shaped magnetic field} 
\citep{2012ApJ...761...40K}.
Indeed, \citet{2007ApJ...659.1404T}
reported that the bipolar outflow is inclined by about $57\degr$
(where $90\degr$ indicates a bipolar outflow on the sky plane).
The inferred inclination makes it possible for the radially pinched
field lines along a given line of sight to produce polarizations
that cancel one another, yielding a less polarized equatorial region
\citep[see Fig.6d of][]{2012ApJ...761...40K}.
{Furthermore, rotation of a cloud introduces a mis-alignment 
to the depolarized strip \citep[see Fig.10d of][]{2012ApJ...761...40K},
which is broadly consistent with the depolarized feature shown in
Figure \ref{fig_polratio}.}

Depolarization can also occur due to high optical depth 
\citep[see Fig.3 of][]{2017MNRAS.472..373Y}, but this is not likely 
the case here since the optical depth is expected to be low along 
this depolarization strip: $\tau \sim 0.08$ even at the highest contour 
level assuming T = 30 K.

In addition, along the depolarized strip, there are several depolarized
clumps, whose sizes are not resolved at our angular resolution, 
as seen in the white in Figure \ref{fig_polratio}.
The clumps are separated by $\sim0.8\arcsec$, which corresponds to 
about 240 au at the distance of Perseus.  These clumps
may indicate relatively more turbulent areas with chaotic magnetic
fields and/or areas with magnetic fields pointing along the line of
sight direction. 
They may even be de-magnetized ``islands'' produced by 
reconnection of sharply pinched magnetic field lines 
\citep[see Fig. 7 of][]{Suriano:2017cm},
although detailed exploration of magnetic reconnection is 
beyond the scope of this paper. 

The highest polarization fractions, reaching levels of up to 40\%, are 
located near the border of the northwest and southeast part of the envelope.
Such high polarization fractions can
occur only by very elongated, aligned dust grains 
{\citep[e.g., axis ratios larger
than 3;][]{1985ApJ...290..211L}.}
Based on the high polarization fraction locations, 
we speculate that mechanical alignment could also contribute
to the polarization in the cavity walls.  In the case of mechanical
alignment, it has recently been suggested that the spinning, minor axis of 
elongated dust grains are
aligned with the mechanical flow \citep{2018ApJ...852..129H}, in
contrast to the classical mechanical alignment \citep{1952MNRAS.112..215G}.
Therefore, the polarization directions could be the same to the dust
grains aligned in poloidal magnetic fields along the bipolar outflow.

{Nevertheless, the largest origin of the high polarization fractions
would be the differential filtering in Stokes maps.}
The continuum structure is more extended in
Stokes I than those in Stokes Q and U, as shown in Figure \ref{fig_iqu}. 
This results in more flux of Stokes I being filtered out, which significantly
contributes to the high polarization fractions, particularly at the edges.

\begin{figure} \plotone{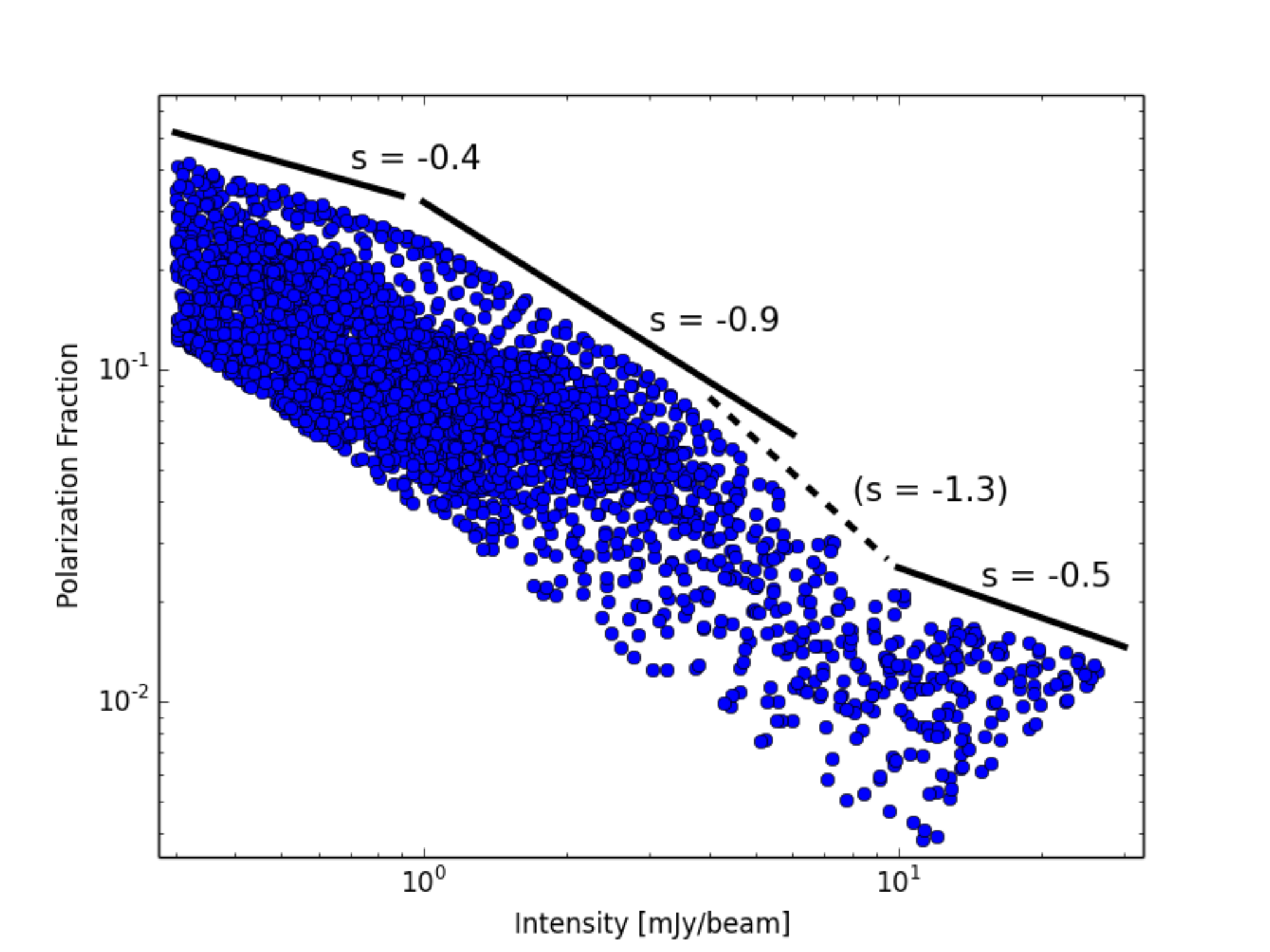} 
\caption{Polarization fraction versus intensity. Each blue circle
represents a pixel value, and the black lines indicate individual power-law
slopes, not fitting results. Data points of intensities greater than $3\sigma$ and
polarization intensities larger than $2\sigma$ have been selected.
\label{fig_pi}} \end{figure}

Figure \ref{fig_pi} shows decreasing polarization
fractions ($P_{frac}$) with Stokes I intensities ($I$) overall.  This 
trend has been reported by many previous polarimetric observations of 
arc-second \citep[e.g.,][]{2006Sci...313..812G,2006ApJ...653.1358K,
2013ApJ...771...71L,2014ApJS..213...13H,2018A&A...616A.139G}
and tens arc-second or coarser angular resolutions 
\citep[e.g.,][]{1996ApJ...470..566D,2016A&A...586A.136P,2018ApJ...861...65S}.
Going toward a central denser region, dust grains get larger causing less
alignment in a magnetic field, optical depth increases, and/or magnetic fields
likely become complicated, which all result in a lower polarization fraction.
{Note that dust grains getting larger, such as above 10 \um\ in disk
conditions, cause less alignment with the magnetic field because
the Larmor precession rate becomes slower than the gas randomization
rate \citep{Hoang:2016fj,2017ApJ...839...56T}.}

Regarding power-law indices of the relationship,
when polarized emission is dominated from the surface of a structure,
the polarization fraction is expected to be inversely proportional 
to the intensity {in the optically thin case} ($P_{frac} \propto I^{-1}$): e.g.,
polarization caused by dust grains aligned by magnetic fields due
to the interstellar radiative torque \citep[RAT,][]{2007MNRAS.378..910L}
mainly around the molecular cloud surface.
Interestingly, Figure \ref{fig_pi} shows multiple
slopes that encompass the limits of the distributions {in a qualitative manner}.  
In the regime fainter than about 1 \mjybeam, the slope is roughly $-0.4$.
This region corresponds to the area from the second lowest contour
to about the fourth contour in the 1 mm continuum map of Figure
\ref{fig_polratio}.  The slope shallower than $-1$ could be interpreted
with additional polarization contributions from these areas (as
addressed above, e.g., mechanical alignment), not limited just on the
structure surface.  
Between $\sim1$ and $\sim10$ \mjybeam, the slope
is close to $-1$, which is indicative that the regions have no
further significant polarization. Approaching to 10
\mjybeam, the slope becomes a little bit steeper than $-1$: $s = -1.3$. 
This can be interpreted as polarization directions changing, resulting
in depolarization.  Indeed, the region is where the field directions switch
from the poloidal to the toroidal pattern.  
For the central region that is brighter than
10 \mjybeam\, the slope is $-0.5$. 
{This shallower slope than $-1$ may be caused by grain alignment 
closer to a central star through RAT, although it would be less efficient
compared to the case of outer fine grains.}
In addition, presumably self-scattering {of emission from large grains} may
also contribute to the polarization fraction at these scales
\citep[e.g.,][]{2017ApJ...844L...5K,2017ApJ...851...55S,2014Natur.514..597S},
{although it may not be dominant in this target.}

\section{Discussion} \label{sec:discuss}
\subsection{Magnetic field strength} \label{subsec:align}

We estimate 
{a very rough} magnetic field strength using the Davis-Chandrasekhar-Fermi
(DCF) method \citep{1951PhRv...81..890D,1953ApJ...118..113C}.
We are aware that the field orientations can be affected by 
outflows and gravitational collapse {near a protostellar system, 
which will likely degrade the accuracy of the DCF method,} but 
quantifying such effects would require more 
detailed dynamical modeling that is beyond the scope of this paper.  
In the DCF technique, the magnetic field strength is estimated based on
the dispersion of magnetic fields with respect to the background
field direction compared to its turbulence in a given density
medium. In other words, a medium at a given density and a turbulence
indicated by a non-thermal linewidth would have a stronger magnetic
field strength when it has a smaller field position angle dispersion:
the plane-of-sky strength of a magnetic field $B_{POS} = Q \sqrt{4
\pi \rho}\, \delta V / \delta \phi \approx 9.3 \sqrt{n(H_2)} \Delta
V / \delta \phi~ [\mu$G], where $Q$, $\rho$, $\delta V$, $\delta
\phi$, and $n(H_2)$ are a factor of order unity, the gas density,
the non-thermal velocity dispersion in \kms, the position angle
dispersion of polarizations, and the molecular hydrogen number
density, respectively
\citep[e.g.,][]{2004ApJ...600..279C,2001ApJ...546..980O}.

\begin{figure} \plottwo{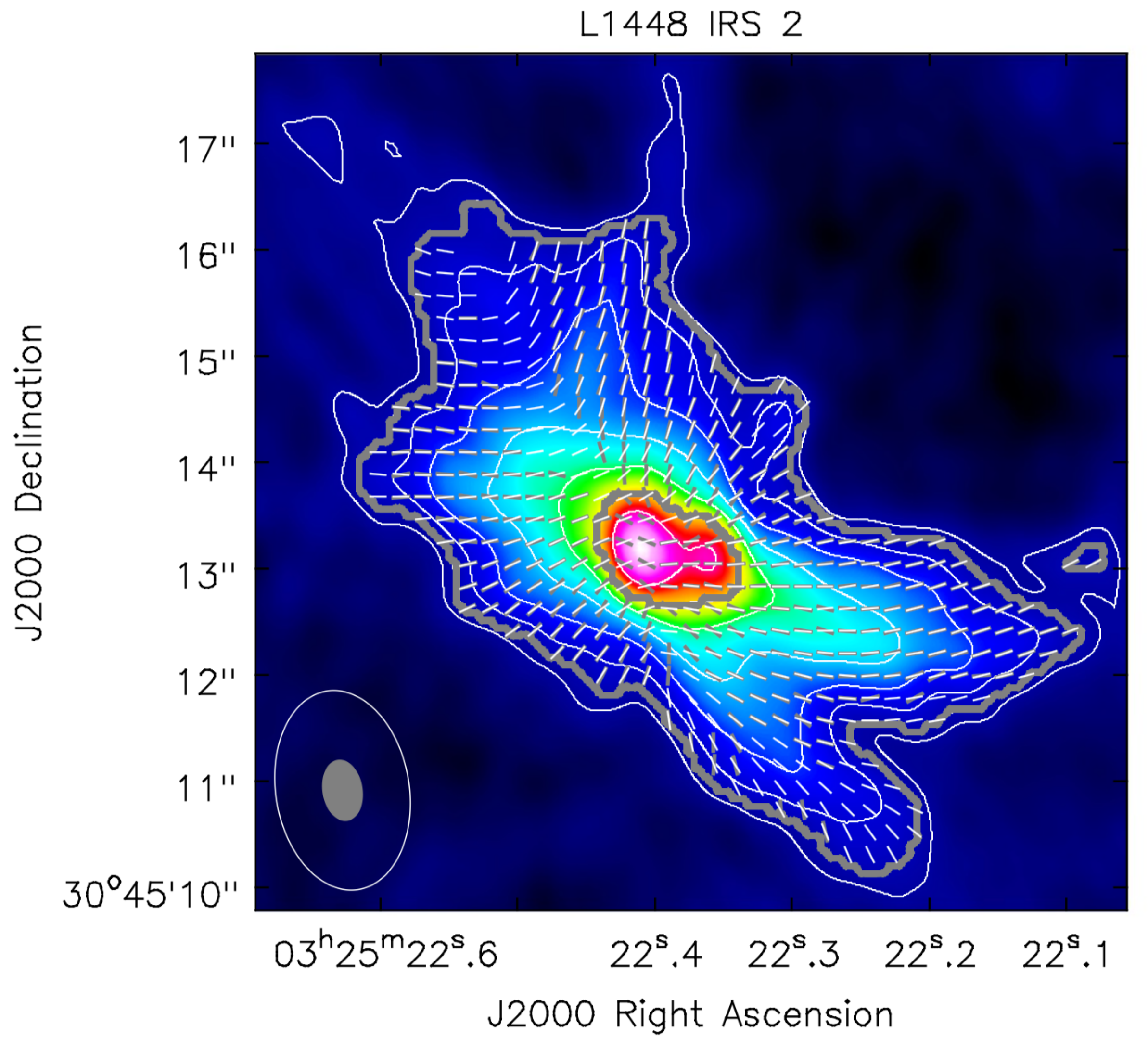}{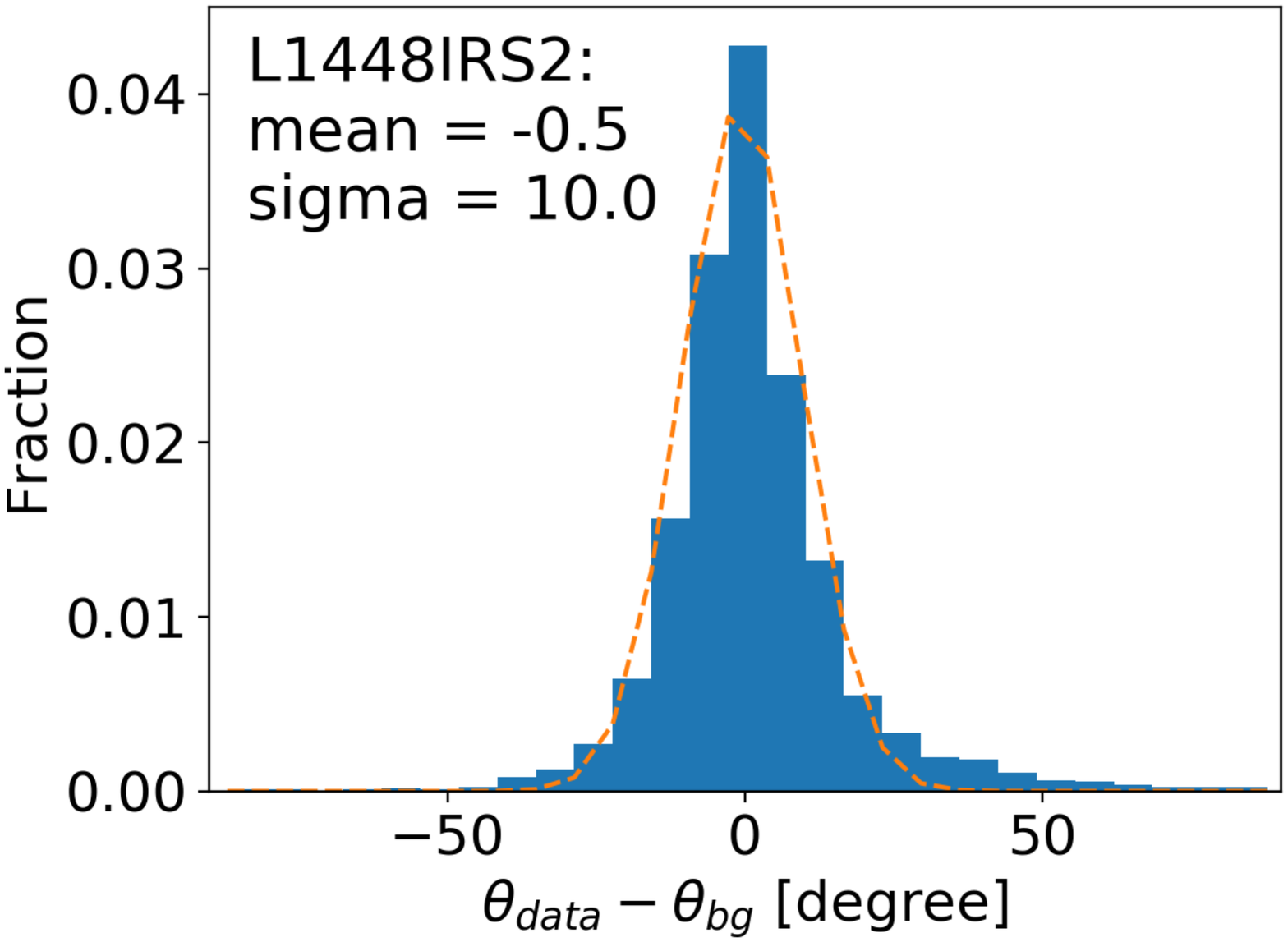}
\caption{Intensity map in color scales.  Magnetic fields of the
original angular resolution are marked in gray and fields smoothed
by nine times larger beams in area are in white.  Refer to the text for
the thick gray contour lines.  In the right is the histogram of
the magnetic field {directions with respect to the background
field.} \label{fig_polsm}} \end{figure}

First, for the background large-scale fields we smoothed the Stokes
Q and U maps with a nine-times larger beam (extended in both major and minor 
axes of the original beam by a factor of three), which is comparable to
a half of the width across the continuum structure. This provides
a reasonable background field morphology \citep{2017ApJ...846..122P}.
As Figure \ref{fig_polsm} shows, the smoothed
fields in white vectors do not show the toroidal switch at the
central region. 
Since we know the central area is confused by the morphological change 
{in scales and with evolution}
and possible polarization contamination from scattering, we only 
apply the DCF method to the areas between intensities of 0.3 and 6.5 
\mjybeam\ (thick gray contours in Fig. \ref{fig_polsm}), for estimating
the magnetic field strength.
The measured dispersion is estimated as $10\degr$ (Fig. \ref{fig_polsm}, 
right). Second,
for estimating the number density of H$_2$ we utilized
the dust continuum. The total continuum flux density of the area between
0.3 and 6.5 \mjybeam, which is $18.2$ squared arc-seconds, is about
91 mJy at 233 GHz.  The total mass is estimated by $M_{T} = F_{\nu}
D^2 / \kappa_\nu B_{\nu} (T_d)$, where $F_{\nu}$, $D$, $\kappa_{\nu}$,
$B_{\nu}$, and $T_d$ are the flux density, distance, mass absorption
coefficient, blackbody radiation intensity, and dust temperature,
respectively.  Using $F_{\nu} = 91$ mJy, $D = 300$ pc, $\kappa_{\nu}
= 0.01~\textrm{cm}^2\,\textrm{g}^{-1}$ at 233 GHz
\citep{1994A&A...291..943O} assuming a gas-to-dust mass ratio of 100,
and $T_d = 30$ K \citep{2009ApJ...696..841K}, 
the total mass is estimated to be 0.08 M$_\sun$.  In
addition, assuming a cylinder with the profile of the continuum
feature, the total volume would be $9\times10^{48}~\textrm{cm}^3$.
Therefore, we derive the volume density $\rho \approx
1.8\times10^{-17}~ \textrm{g}\, \textrm{cm}^{-3}$, which
corresponds to $n(H_2) \approx 5.3\times10^6$ cm$^{-3}$. We do
not have an observational non-thermal linewidth, but it may be
reasonable to adopt the trans-sonic velocity at 30 K: $\sim 0.3$
\kms. These values result in the magnetic field strength in the
plane of the sky of about 640 $\mu$G, with the relationship following: 
\begin{equation} B_{POS}~ \approx
640 ~\mu\textrm{G}~ \Big( \frac{n(H_2)}{5.3\times10^6 \textrm{cm}^{-3}}
\Big)^{0.5} \Big(\frac{\Delta V}{0.3~\textrm{km/s}} \Big) \Big(
\frac{10\degr}{\delta \phi} \Big).  \end{equation}

Furthermore, we estimate the magnetic braking time scale of the
presumed disk structure at the center, when the rotation velocity
decreases by a half \citep[e.g.,][]{1994ApJ...432..720B}.
The Alfv\'en speed follows the relationship, 
\begin{equation} v_A = \frac{B}{\sqrt{4 \pi
\rho}} = 0.43~\textrm{km/s}~ \Big( \frac{B}{640~\mu \textrm{G}} \Big)
\Big( \frac{1.8\times10^{-17}~\textrm{g/cm}^3}{\rho} \Big)^{0.5}.
\end{equation} 
In addition, the central mass surrounded by the inner
thick gray contour in Figure \ref{fig_polsm} is estimated as 0.05
M$_\sun$ based on the total flux density of 55 mJy. 
{This mass is rather uncertain: it could be overestimated because
the central region is warmer than the outer region and could be
underestimated because the very central region ($<$ a few au in radius) 
would be optically thick even in millimeter wavelengths.}
{On the other hand,} the central rotating structure could be much 
smaller than the inner
region considered here. The same mass beyond the central area is
extended up to the intensity of $\sim1.1$ \mjybeam, which is
about $0.5\arcsec$ (150 au) away. When the Alfv\'en wave reaches
this point, the rotating mass tied up by the magnetic field is doubled so
the rotation velocity becomes a half assuming angular momentum
conservation.  This timescale is calculated to be $\sim1700$ years.
Note that this is 
much shorter than the typical age of Class 0 YSOs, which is several
thousands years. Even further, when regarding the canonical accretion
rate of Class 0 YSOs $\sim10^{-6}$ M$_\sun$ year$^{-1}$
\citep[e.g.,][]{1977ApJ...214..488S,2014prpl.conf..195D}, 
the magnetic braking effect,
which slows down 0.05 M$_\sun$ in 1700 years, dominates the system.
Taken at face value, the estimated field strength is high 
enough for the magnetic field to brake the disk rotation efficiently.
However, as we mentioned earlier, the polarization orientations on 
the several tens au scale are indicative of a disk. If true, the 
existence of a relatively large disk in the presence of a strong 
inferred magnetic field would point to a decoupling of the field 
from the bulk disk material, most likely through non-ideal MHD 
effects, which become more important at higher densities
{\citep[e.g.,][]{2010ApJ...718L..58I,2011ApJ...733...54K, Dapp:2012fo, 
2015ApJ...801..117T, {2015ApJ...810L..26T}, 2018MNRAS.473.4868Z}.}

\section{Conclusion}

We have detected a well-ordered polarization pattern
toward the Class 0 YSO L1448 IRS 2, whose inferred magnetic field
presents the clearest hourglass morphology to date on 100 au scales:
poloidal in the outer regions and rapidly switching to toroidal
in the inner region.  This can be interpreted as a toroidal magnetic
field wrapped up by a rotating (disk) structure or by a self-scattered
polarization pattern due to large grains in an inclined disk: 
either case supports a rotationally dominant structure.  
Future high resolution molecular line observations are needed to 
investigate whether there is a rotationally supported disk.

We found four regimes with different slopes in the relationship
between polarization fractions and intensities, which provide
interesting constraints on grain alignment mechanisms.
In addition, we detected a clumpy depolarization strip, which
is indicative of magnetically channelled protostellar accretion flows 
which drag the field lines into a radially pinched configuration that, 
when combined with inclination effects, lowers the degree of polarization. 

Finally, we estimated the plane-of-sky magnetic field strength using
the DCF technique and found that magnetic braking should be
very efficient in the system, which is inconsistent with the strong
hints of a central disk, protobinary, and  observations of rotation.
Therefore, the magnetic braking
catastrophe based on {simple} ideal MHD simulations may not be so disastrous,
{at least in this source.
Our observations emphasize that} non-ideal MHD effects 
(and possibly turbulence) should be taken into account, in order to fully
understand the formation of disks at the early protostellar systems, 


\acknowledgments
We are grateful to ALMA staff for their dedicated work and an anonymous referee
for helpful comments.  W.K. thanks
Thiem Hoang for fruitful discussions on grain alignments and
polarization mechanisms.  W.K. was supported by Basic Science
Research Program through the National Research Foundation of Korea
(NRF-2016R1C1B2013642).  Z.-Y.L. is supported in part by NSF AST-1313083 
and AST-1716259 and NASA NNX14AB38G and 80NSSC18K1095. 
This paper makes use of the following ALMA
data: ADS/JAO.ALMA\#2016.1.00604.S, ADS/JAO.ALMA\#2013.1.00031.S.
ALMA is a partnership of ESO (representing its member states), NSF
(USA) and NINS (Japan), together with NRC (Canada), MOST and ASIAA
(Taiwan), and KASI (Republic of Korea), in cooperation with the
Republic of Chile. The Joint ALMA Observatory is operated by ESO,
AUI/NRAO and NAOJ.

%

\vspace{5mm} \facilities{ALMA}

\bibliographystyle{apj} \bibliography{class0pol}



\end{document}